\documentclass{JHEP} 
\def\dfrac#1#2{{\displaystyle\frac{#1}{#2}}}
\author{Alexander A.~Chernitskii\\
St.Petersburg Electrotechnical University,\\
Prof. Popov str. 5, St.Petersburg 197376,  Russia\\
E-mail: \email{aa@cher.etu.spb.ru}}
\title{{ Light beams distortion\\ in nonlinear electrodynamics}}
\abstract{
We obtain the characteristic equation for the nonlinear
Born-Infeld electrodynamics. This equation has the form
of the characteristic equation for the linear electrodynamics
in some effective Riemann space.
The effective metric include the energy-momentum tensor components
of electromagnetic field.
We study a distortion of light beams by the action of some
distant solitons.  This distortion corresponds
to attraction with the solitons and
looks like the gravitational distortion.
}

\preprint{\jhep{11}{1998}{015}\\ \hepth{9809175}}
\received{October 9, 1998}
\accepted{November 19, 1998}

\begin{document}
\mathsurround=2pt
%\maketitle

\section{Introduction}

The dynamics of solitons looks like the dynamics of particles.
Specifically, the electromagnetic solitons behave like
gravitating particles if we consider the second order by
a field of distant solitons \cite{I1992,I1995_GR14}.
In the first order of this case we have the Lorentz force.

At the moment we reason that
the nonlinear electrodynamics with singularities
discussed in the article \cite{I1998_HPA}
is best suited to this analogues. In this article we had considered a
modernized Born-Infeld action that include a singular part with
\mbox{$\delta$-functions}.
This is equivalent also to introducing boundary conditions at
points. These points behave like point charged particles.
In this case we have the space with a non-trivial topology.

A subject of the present article is a study
of light beams distortion as nonlinear effect of the model.

\section{Field equations outside of the singularities}

In this article we consider the field equations outside of
the singularities that is the purely Born-Infeld model.
Thus the system of equations may by written
in the following form:
\begin{equation}
\label{Eq:Max1}
\left\{
\begin{array}{rcl}
{\rm div}{\bf B} &=& 0\\[7pt]
{\rm div}{\bf D} &=& 0
\end{array}
\right.
\quad,\qquad\left\{
\begin{array}{rcl}
\phantom{-}\partial_0 \, {\bf B}  {}+{}  {\rm rot}{\bf E}  &=&
0\\[7pt]
-\partial_0 \, {\bf D}  {}+{}  {\rm rot}{\bf H}  &=& 0
\end{array}
\right.
\quad,
\end{equation}
where
\begin{eqnarray}
\left\{
\begin{array}{ccl}
{\bf D} &=& \displaystyle\frac{1}{{\cal L}}\,({\bf E}  {}+{}
\alpha^2\,{\cal J} {\bf B})\\[9pt]
{\bf H} &=& \displaystyle\frac{1}{{\cal L}}\,({\bf B}  {}-{}
\alpha^2\,{\cal J} {\bf E})
\end{array}
\right.\quad,\qquad
\left\{
\begin{array}{ccl}
{\bf E} &=& \displaystyle\frac{1}{{\overline {\cal L}}}\,({\bf D}
{}-{}  \alpha^2\,{\overline{\cal J}} {\bf H})\\[9pt]
{\bf B} &=& \displaystyle\frac{1}{{\overline {\cal L}}}\,({\bf H}
{}+{}  \alpha^2\,{\overline{\cal J}} {\bf D})
\end{array}
\right.
\quad,
\label{MatEq}
\end{eqnarray}
\begin{equation}
\begin{array}{l}
{\cal L} {\,}={\,}
\sqrt{\vphantom{{\overline{\cal J}}^2}|\,1 {}-{}  \alpha^2\,{\cal I}
{}-{}  \alpha^4\,{\cal J}^2\,|}
\quad,\qquad
{\overline {\cal L}}  {}\equiv{}
\sqrt{|1 {}-{} \alpha^2 \,{\overline{\cal I}} {}-{}
\alpha^4 \,{\overline{\cal J}}^2 |}
\quad,\\[9pt]
{\cal I} {\,}={\,} {\bf E}^2 {}-{} {\bf B}^2
\quad,\qquad
{\overline{\cal I}} {\,}={\,} {\bf H}^2 {}-{} {\bf D}^2
\quad,\qquad
{\cal J} {\,}={\,} {\overline{\cal J}} {\,}={\,}
{\bf E} {}\cdot{} {\bf B} {\,}={\,} {\bf H} {}\cdot{} {\bf D}
\quad.
\\[7pt]
\end{array}
\end{equation}
Here we suppose that the components of metric are
$g_{00} {}={} -1$, $g_{0i} {}={} 0$.

We can use only the second block of eqs. (\ref{Eq:Max1})
for a problem with initial conditions.
For a cartesian coordinate system we can write the following
system:
\begin{equation}
\label{Eq:Max3}
\left\{
\begin{array}{rcl}
\displaystyle\frac{\partial B_i}{\partial x^0}  {}+{}
\varepsilon_{ijk}\, \displaystyle\frac{\partial E_k}{\partial x^j}
&=& 0\\[11pt]
- C_{{\rm DE}}^{ij}\,
\displaystyle\frac{\partial E_j}{\partial x^0}  {}-{}
C_{{\rm DB}}^{ij}\,
\displaystyle\frac{\partial B_j}{\partial x^0} {}+{}
\varepsilon_{ijk}\,
C_{{\rm HE}}^{kl}\,
\displaystyle\frac{\partial E_l}{\partial x^j} {}+{}
\varepsilon_{ijk}\,
C_{{\rm HB}}^{kl}\,
\displaystyle\frac{\partial B_l}{\partial x^j}
&=& 0
\end{array}
\right.
\quad,
\end{equation}
where
\begin{equation}
\label{DEHB}
\left\{
\begin{array}{rcl}
C_{{\rm DE}}^{ij} &=&
\delta^{ij} {}+{} \alpha^2 \left(D_i\,D_j {}+{} B_i\,B_j \right)
\\[5pt]
C_{{\rm DB}}^{ij} &=& \alpha^2
\left( B_i\,E_j {}-{} D_i\,H_j \right)\\[5pt]
C_{{\rm HB}}^{kl} &=&
\delta_{kl} {}-{} \alpha^2 \left(H_k\,H_l {}+{} E_k\,E_l \right)
\\[5pt]
C_{{\rm HE}}^{kl} &=& \alpha^2
\left( D_l\,H_k {}-{} B_l\,E_k \right)
\end{array}
\right.
\end{equation}
and the Latin indexes take the values $1,\,2,\,3$.

Let us write the system of equations (\ref{Eq:Max3}) in the following
matrix form:
\begin{equation}
\label{Eq:Max3m}
Q^\mu\,\displaystyle\frac{\partial}{\partial x^\mu}\left(
\matrix{{\bf E}\cr {\bf B}}\right) {}={} 0
\quad,
\end{equation}
where $(6 {}\times{} 6)$ matrix $Q^\mu {}={} Q^\mu ({\bf E},{\bf B})$
may by easy obtained from eqs. (\ref{Eq:Max3}), (\ref{DEHB}).

\section{Characteristic equation}

The characteristic equation for the system of such type
((\ref{Eq:Max3}) or (\ref{Eq:Max3m}))
is given by the following relation \cite{CurantHilbert}:
\begin{equation}
\label{Eq:Char}
\det (Q^\mu\,k_\mu) {}={} 0
\quad,
\end{equation}
where
\begin{equation}
\label{Def:k}
k_\mu {}\equiv{} \displaystyle\frac{\partial {\cal S}}{\partial x^\mu}
\end{equation}
and ${\cal S}\bigl( x \bigr) {}={} 0$
is the equation of a characteristic surface.

For the system (\ref{Eq:Max3}-\ref{Eq:Max3m}) we have the following
expressions:
\begin{eqnarray}
\label{Expr:Det1}
\det (Q^\mu\,k_\mu) &=&
\displaystyle\frac{\left( k_0 \right)^2}{{\cal L}^2}\,
\left[\left( h^{\mu\nu} {}-{}
\alpha^2\,{\cal F}^{\mu\rho}\,{\cal F}^{\nu}_{.\rho} \right)
\,k_\mu\,k_\nu\right]^2
\\[7pt]
\label{Expr:Det2}
&=&
\left( k_0 \right)^2\,{\cal L}^2\,\left[\left( h^{\mu\nu} {}-{}
\alpha^2\,f^{\mu\rho}\,f^\nu_{.\rho} \right)\,k_\mu\,k_\nu\right]^2
\\[11pt]
\label{Expr:Det3}
&=&
\left( k_0 \right)^2\,\left[\left( h^{\mu\nu} {}-{}
\alpha^2\,{\overline{T}}^{\mu\nu} \right)\,k_\mu\,k_\nu\right]^2
\quad,
\end{eqnarray}
where\quad
$h_{11} {}={} h_{22} {}={} h_{33} {}={} -h_{00} {}={} 1$\quad,\qquad
$h_{\mu\nu}  {}={}  0$ \quad for\quad $\mu\neq\nu$\quad;\\
\phantom{where}\quad
the Greek indexes take the values $0,\,1,\,2,\,3$\quad,
\begin{eqnarray}
\label{Def:cF}
& &{\cal F}^{\mu\nu} {}={}
-\frac{1}{2}\,\varepsilon^{\mu\nu\sigma\rho}\,F_{\sigma\rho}
\quad,\qquad
F_{i0} {}={}E_i
\quad,\qquad
F_{ij} {}={} \varepsilon_{0ijl}\,B^l
\quad,\\[5pt]
& &
\label{Def:f}
f^{\mu\nu} {}={}
\frac{1}{\alpha^2}\,\frac{\partial{\cal L}}{\partial F_{\mu\nu}}
\quad,\qquad
f^{0i} {}={} D^i
\quad,\qquad
f^{ij} {}={} -\varepsilon^{0ijl}\,H_l
\quad,\\[5pt]
& &{\overline{T}}^{\mu\nu} {}\equiv{} f^{\mu\rho}\,F^\nu_{.\rho}  {}-{}
\displaystyle\frac{1}{\alpha^2}\,\left({\cal L} {}-{} 1\right)\,h^{\mu\nu}
\quad,
\label{Def:TEI}
\end{eqnarray}
$\varepsilon_{0123} {}={} 1$, $\varepsilon^{0123} {}={} -1$,
${\overline{T}}^{\mu\nu}$ are the metric energy-momentum tensor
components \cite{I1998_HPA}.

Suppose $k_0 {}\neq{} 0$. Then, we have the following
very interesting form of the characteristic equation
generalized to the case of any metric:
\begin{equation}
\label{Eq:Char2}
\left( g^{\mu\nu} {}-{}
\alpha^2\,{\overline{T}}^{\mu\nu} \right)\,k_\mu\,k_\nu
{}={} 0
\quad.
\end{equation}

\section{Distortion of beams}

From definition (\ref{Def:k}) it follows that
$\partial_\mu k_\nu {}-{} \partial_\nu k_\mu {}={} 0$; whence we get
(see also \cite{Witham})
\begin{equation}
\label{Eq:beem}
\frac{{\rm d} {\bf k}}{{\rm d} x^0} {\,}={\,}
-\mbox{{\bf $\nabla$}}\,W({\bf k},x)
\quad,
\end{equation}
where
\begin{equation}
\label{Rel:Disp}
-k_0 {\,}\equiv{\,} \omega {\,}={\,} W({\bf k},x)
\end{equation}
is a form of the characteristic equation, that may be obtained
from the form of type (\ref{Eq:Char2}).
Here the full derivative on time is defined as
\begin{equation}
\frac{{\rm d} k_i}{{\rm d} x^0} {\,}\equiv{\,}
\frac{\partial k_i}{\partial x^0}
{}+{}
V^j\,\frac{\partial k_i}{\partial x^j}
\quad,
\end{equation}
where $V^j {}\equiv{} \dfrac{\partial W}{\partial k_j}$
are the components of group velocity.

The vector ${\bf k}$ is normal to the two-dimensional surface
${\cal S} (x^0,\,{\bf x} ) {}={} 0$ in any moment of time. Thus
the equation (\ref{Eq:beem}) may be called the beam one.

From the characteristic equation (\ref{Eq:Char2}) it follows
the expression for the function W.
\begin{equation}
\label{Expr:Ws}
W {}={}
\frac{\alpha^2\,{\overline{T}}^{0i}\,k_i {}\pm{}
\sqrt{\left(\alpha^2\,{\overline{T}}^{0i}\,k_i\right)^2 {}+{}
\left(1 {}+{} \alpha^2\,{\overline{T}}^{00} \right)
\left({\bf k}^2 {}-{}
\alpha^2\,{\overline{T}}^{ij}\,k_i\,k_j\right)}}
{1 {}+{} \alpha^2\,{\overline{T}}^{00}}
\quad.
\end{equation}
Let us express the function $W$ (\ref{Expr:Ws})
as a power series in $\alpha^2$.
This is correspond to the small field approximation
$|\alpha\,{\bf E}| {}\ll{} 1,\,|\alpha\,{\bf B}| {}\ll{} 1$.
Here we disregard the terms with $\alpha^4$ and higher ones.
Thus we obtain
\begin{eqnarray}
\label{Expr:Wser}
W &=&
\alpha^2\,{\overline{T}}^{0i}\,k_i {\,}\pm{\,} |{\bf k}| \left[
1 {}-{} \frac{\alpha^2}{2} \left( {\overline{T}}^{00} {}+{}
{\overline{T}}^{ij}\,\frac{k_i\,k_j}{{\bf k}^2} \right)\right]
\quad.
\end{eqnarray}

Substituting (\ref{Def:TEI}) for ${\overline{T}}^{\mu\nu}$ in
(\ref{Expr:Wser}),
using (\ref{Def:cF}), (\ref{Def:f}),
and ignoring $\alpha^4$ terms,  we get
\begin{eqnarray}
\label{Expr:WserEB}
W &=&\alpha^2
\left( {\bf E} {}\times{} {\bf B} \right) {}\cdot{} {\bf k} {\;}\pm{\,}
\left\{ |{\bf k}| {}-{} \frac{\alpha^2}{2\,|{\bf k}|}
\left[\left({\bf E} {}\times{} {\bf k}  \right)^2  {}+{}
\left({\bf B} {}\times{} {\bf k}  \right)^2
\right]\right\}
\end{eqnarray}

Now we consider the propagation of a
low-amplitude high-frequency electromagnetic wave
in the presence of some given field.
We can consider a field of some distant solitons as the given field.

The model
(\ref{Eq:Max1}),(\ref{MatEq}) has the following plane wave solution:
\begin{equation}
\label{Sol:PlaneW}
\tilde{{\bf E}} {}={}
\frac{1}{2}\,
\left({\bf u}\,{\rm e}^{{\rm i} \Theta}
{}+{} {\bf u}^{*}\,{\rm e}^{-{\rm i} \Theta}
\right)
\quad,\qquad
\tilde{{\bf B}} {}={}
\frac{1}{2}\,
\bar{\bf k} {}\times{}
\left({\bf u}\,{\rm e}^{{\rm i} \Theta}
{}+{} {\bf u}^{*}\,{\rm e}^{-{\rm i} \Theta}
\right)
\quad,
\end{equation}
where
$\Theta {}={} \omega\,x^0 {}-{} {\bf k} {}\cdot{} {\bf x}$,
$\omega^2 {}={} {\bf k}^2$,
 ${\bf u}$ is a complex vector amplitude that
${\bf u} {}\cdot{} {\bf k} {}={} 0$,
$\bar{\bf k} {}\equiv{} {\bf k}/|{\bf k}|$.
The field configuration (\ref{Sol:PlaneW}) describe the plane wave
with any polarization. We find a solution of equation
(\ref{Eq:Max3m}) in the following form:
\begin{equation}
\label{Sol:Sum}
\left(\matrix{{\bf E}\cr {\bf B}}\right) {}={}
\left(\matrix{\tilde{{\bf E}}\cr\tilde{{\bf B}}}\right) {}+{}
\left(\matrix{\underline{{\bf E}}\cr
\underline{{\bf B}}}\right)
\quad,
\end{equation}
where $\underline{{\bf E}},\,\underline{{\bf B}}$ is a given field and
we suppose that the phase $\Theta (x)$ of the solution (\ref{Sol:PlaneW})
is a some unknown function and
$k_\mu {}\equiv{} \dfrac{\partial \Theta}{\partial x^\mu}$
($\omega {}\equiv{} -k_0$).\\[11pt]
We suppose also that
$\left(\matrix{\tilde{{\bf E}}\cr \tilde{{\bf B}}}\right)
 {}\ll{}
\left(\matrix{\underline{{\bf E}}\cr \underline{{\bf B}}}\right)$
and
$\omega {}\gg{}
{\displaystyle\left.\left|
\frac{\partial \underline{{\bf E}}}{\partial x^\mu}\right|\right/
\left|\underline{{\bf E}}\right|}$.\\[11pt]
Then substituting (\ref{Sol:Sum}) in (\ref{Eq:Max3m}),
we get\quad
$Q^\mu\,k_\mu\left(
\matrix{{\bf u}\cr \bar{\bf k} {}\times{} {\bf u}}\right) {}={} 0
$,
where $Q^\mu {}={} Q^\mu (\underline{{\bf E}},\underline{{\bf B}})$.\\[11pt]
This equation has non-trivial solutions when $\det (Q^\mu\,k_\mu) {}={} 0$.
Thus we have the characteristic equation as dispersion relation.
If in addition
\mbox{$|\alpha\,\underline{{\bf E}}| {}\ll{} 1$},
\mbox{$|\alpha\,\underline{{\bf B}}| {}\ll{} 1$}, then we have the
dispersion function $W {}={} W({\bf k},\underline{{\bf E}},
\underline{{\bf B}})$
in the form (\ref{Expr:Wser}) or (\ref{Expr:WserEB}).

The characteristic equation (\ref{Eq:Char2}) or dispersion
relation $\omega {}={} W({\bf k},\underline{{\bf E}},\underline{{\bf B}})$
define a family of surfaces
\mbox{${\cal S} {}={} \Theta {}={} {\rm const} $}. We can introduce
a curvilinear coordinates $\{ x^{\prime i} \}$ that
$x^{\prime 1}$ is perpendicular to these surfaces.
Then $k^\prime_2 {}={} k^\prime_3 {}={} 0$.
Let us define the direction of the axis $x^{\prime 1}$ that
$k^\prime_1 {}={} |{\bf k}|$.

The phase velocity ${\bf v}$ of wave satisfy the
equation $W {}-{} {\bf k} {}\cdot{} {\bf v} {}={} 0$.
Using (\ref{Expr:WserEB}), for the coordinate system
$\{x^0,{\bf x}^\prime\}$
we obtain the following values of the phase velocity:
\begin{equation}
\begin{array}{rcl}
v_+ &=&  1 {}-{} \dfrac{\alpha^2}{2}
\left[\left( E^\prime_2 {}-{}
B^\prime_3 \right)^2 {}+{}
\left( E^\prime_3 {}+{}
B^\prime_2 \right)^2\right] \\[7pt]
v_- &=&  1 {}-{} \dfrac{\alpha^2}{2}
\left[\left( E^\prime_2 {}+{}
B^\prime_3 \right)^2 {}+{}
\left( E^\prime_3 {}-{}
B^\prime_2 \right)^2\right]
\end{array}
\quad,
\end{equation}
where $|{\bf v}|{}={} v_+ $ for the wave that propagate in
the positive direction of the axis $x^{\prime 1}$,
$|{\bf v}| {}={} v_-$ for the wave that propagate in
the opposite direction.

As we see, the two magnitudes of the phase velocity less then
the speed of light. These magnitudes decrease when the
given field increase.
Thus according to the beam equation (\ref{Eq:beem})
we have the light beam distortion corresponding
to attraction with distant solitons.

\section{Conclusion}

We have obtained the characteristic equation (\ref{Eq:Char2})
for the Born-Infeld
nonlinear electrodynamics. This equation has the form of
the characteristic equation for the linear electrodynamics
$\bar{g}^{\mu\nu}\,k_\mu\,k_\nu {}={} 0$
in some effective Riemann space.
The effective metric\break
include the energy-momentum tensor components
of electromagnetic field\break
\mbox{$\bar{g}^{\mu\nu} {}={} g^{\mu\nu} {}-{}
\alpha^2\,{\overline{T}}^{\mu\nu}$}. According to this equation we
have the distortion of light beams that
corresponds to attraction with distant solitons.
This looks like the gravitational distortion.

This property of light beams for the nonlinear model (\ref{Eq:Max1}),
(\ref{MatEq}) are in general agreement with the results obtained
previously for motion of solitons \cite{I1995_GR14}.
But the motion of the solitons with singularities \cite{I1998_HPA}
need further consideration.

\end{document}